\documentclass[10pt,conference]{IEEEtran}
\IEEEoverridecommandlockouts
\usepackage{cite}
\usepackage{amsmath,amssymb,amsfonts}
\usepackage{algorithmic}
\usepackage{graphicx}
\usepackage{textcomp}
\usepackage{xcolor}
\usepackage{balance}

\usepackage{cite}
\usepackage{amsmath,amssymb,amsfonts}
\usepackage{algorithmic}
\usepackage{graphicx}
\usepackage{enumerate}
\usepackage{textcomp}
\usepackage{xcolor}
\usepackage{tcolorbox}
\usepackage{listings}
\usepackage[utf8]{inputenc}
\usepackage{hyperref}
\usepackage{booktabs}
\usepackage{csquotes}
\usepackage{subcaption}
\usepackage{caption}

\def\BibTeX{{\rm B\kern-.05em{\sc i\kern-.025em b}\kern-.08em
    T\kern-.1667em\lower.7ex\hbox{E}\kern-.125emX}}

\graphicspath{{figs/}}

\definecolor{codegreen}{rgb}{0,0.6,0}
\definecolor{codeblue}{rgb}{0,0,0.6}
\definecolor{codegray}{rgb}{0.5,0.5,0.5}
\definecolor{codepurple}{rgb}{0.58,0,0.82}
\definecolor{backcolour}{rgb}{0.95,0.95,0.92}

\def\isdraft{1}

\newcommand{\mynote}[3]{\fbox{\bfseries\sffamily\scriptsize #1}{\small\textsf{\emph{\color{#3}{ #2}}}}}

\if\isdraft1
\newcommand{\petrillo}[1]{\mynote{Petrillo}{#1}{red}}
\newcommand{\jalves}[1]{\mynote{Jalves}{#1}{codegreen}}
\newcommand{\todo}[1]{\mynote{TODO}{#1}{codepurple}}
\newcommand{\argument}[1]{\mynote{Argument:}{#1}{codeblue}}

\else
\newcommand{\petrillo}[1]{}
\newcommand{\jalves}[1]{}
\newcommand{\todo}[1]{}
\newcommand{\argument}[1]{}
\fi

\lstdefinestyle{mystyle}{
	backgroundcolor=\color{backcolour},   
	commentstyle=\color{codegreen},
	keywordstyle=\color{magenta},
	numberstyle=\tiny\color{codegray},
	stringstyle=\color{codepurple},
	basicstyle=\ttfamily\footnotesize,
	breakatwhitespace=false,         
	breaklines=true,                 
	captionpos=b,                    
	keepspaces=true,                 
	numbers=left,                    
	numbersep=5pt,                  
	showspaces=false,                
	showstringspaces=false,
	showtabs=false,                  
	tabsize=2
}

\newtcolorbox[auto counter]{hypothesis}[1]
{
	coltitle=black,
	colframe=gray!25,
	arc = 0.5mm,
	title=\textbf{Hypothesis~\thetcbcounter:} (#1)
}

\begin{document}

\title{Towards improving architectural diagram consistency using system descriptors}


	
	

\author{

	\IEEEauthorblockN{Jalves Nicacio}
	\IEEEauthorblockA{
		\textit{Université du Québec à Chicoutimi} \\
		Chicoutimi, Québec, Canada \\
		jalves.nicacio@ifal.edu.br
	}
	
	\and
	
	\IEEEauthorblockN{Fabio Petrillo}
	\IEEEauthorblockA{
		\textit{Université du Québec à Chicoutimi} \\
		Chicoutimi, Québec, Canada \\
		fabio@petrillo.com
	}
}

\maketitle

\begin{abstract}
Communication between practitioners is essential for the system's quality in the DevOps context. 
To improve this communication, practitioners often use informal diagrams to represent the components of a system.
However, as systems evolve, it is a challenge to synchronize diagrams with production environments consistently. Hence, the inconsistency of architectural diagrams can affect communication between practitioner and their understanding of systems.
In this paper, we propose the use of system descriptors to improve deployment diagram consistency. We state two main hypotheses: (1) if an architectural diagram is generated from a valid system descriptor, then the diagram is consistent; (2) if a valid system descriptor is generated from an architectural diagram, then the diagram is consistent.
We report a case study to explore our hypotheses. Furthermore, we constructed a system descriptor from the Netflix deployment diagram, and we applied our tool to generate a new architectural diagram. Finally, we compare the original and generated diagrams to evaluate our proposal.
Our case study shows all Docker compose description elements can be graphically represented in the generated architectural diagram, and the generated diagram does not present inconsistent aspects of the original diagram. Thus, our preliminary results lead to further evaluation in controlled and empirical experiments to test our hypotheses.
\end{abstract}


\section{Introduction}
\label{sec:introduction}

In the software architecture modelling process, several abstraction levels can represent systems across different resources, such as text or diagrams. Architectural diagrams are useful for communication and understanding of systems, as they motivate a more active discussion among participants as facilitating the memorization of details about systems \cite{Jolak2020}. In this sense, architectural diagrams play a relevant role in facilitating system comprehension. However, architectural diagrams (as UML deployment diagrams) usually remain schematic and disassociated from production reality. Furthermore, architectural diagrams do not follow the evolution of the systems, and there are challenges for engineers to synchronize the diagrams with the system in production so that they always reflect the current state of the system. 

Informally, we analysed architectural diagrams from technical blog posts of Amazon, LinkedIn, or Netflix. 
The authors of these posts are software engineers who designed and presented architectural diagrams intending to communicate information about the systems they described in their articles.
We noticed that engineers use general-purpose notations to create models in tools such as Visio or previous draw.io\footnote{The tool is currently called Diagrams.net}. However, the use of general-purpose tools to draw architectural diagrams can lead mistakes and inconsistencies \cite{Brown2020}.  

In the context of systems modelling, Litvak \textit{et al.} \cite{Litvak2003} states that a consistency problem can arise because more than one diagram can describe the same aspects of the model. Furthermore, diagrams are not updated as the system evolves, and there is no guarantee that diagrams are consistent with the system in production. 



In this paper, we propose the use of system descriptors towards improving architectural diagrams consistency. System descriptors are scripts used in DevOps for automation, standardization, and infrastructure management in production environments. We can mention Puppet, Chef, Docker Compose, and Kubernetes Pods as some of the most well-known system descriptors. The novelty of this research is the use of system descriptors to validate architectural diagrams consistency. 

Our main contributions are:
\begin{itemize}
    \item We state two hypotheses: (1) if an architectural diagram is generated from a valid system descriptor then the diagram is consistent; (2) if a valid system descriptor is generated from an architectural diagram then the diagram is consistent.
    \item  A tool to generate architectural diagrams from docker-compose files. According to a meta-model, the tool analyzes a docker-compose file and then transforms its tags into visual elements of the corresponding diagram.
\end{itemize}

Given the widespread use of system descriptors in modern system architectures, our approach can potentially impact how to produce and validate architecture diagrams in the future.

The remainder of this paper is organized as follows:
Section \ref{sec:motivation} presents the motivation for this work.
In Section \ref{sec:approach} we present concepts behind our approach, our hypotheses are in Section \ref{sec:hypotheses}, and Section \ref{sec:Transformation-function} describe the transformation function.
In Section \ref{sec:case-study}, we describe a study carried out with a tool that generates architectural diagrams from Docker-Compose files.
In Section \ref{sec:discussion}, we discuss a preliminary evaluation.
Finally, in Section \ref{sec:conclusion}, we draw some conclusions and outline directions for future work.
%
%
\section{Are architectural diagrams inconsistent?}
\label{sec:motivation}

In a natural language context, consistency is the state or condition of always happening (or behaving) in the same way\cite{consistency-cambridge}. In the context of systems modelling, Litvak \textit{et al.} \cite{Litvak2003} states that a consistency problem can arise because more than one diagram can describe the same aspects of the model. Furthermore, diagrams are not updated as the system evolves. 

\begin{figure}[h]
    \centering
    \includegraphics[width=\columnwidth]{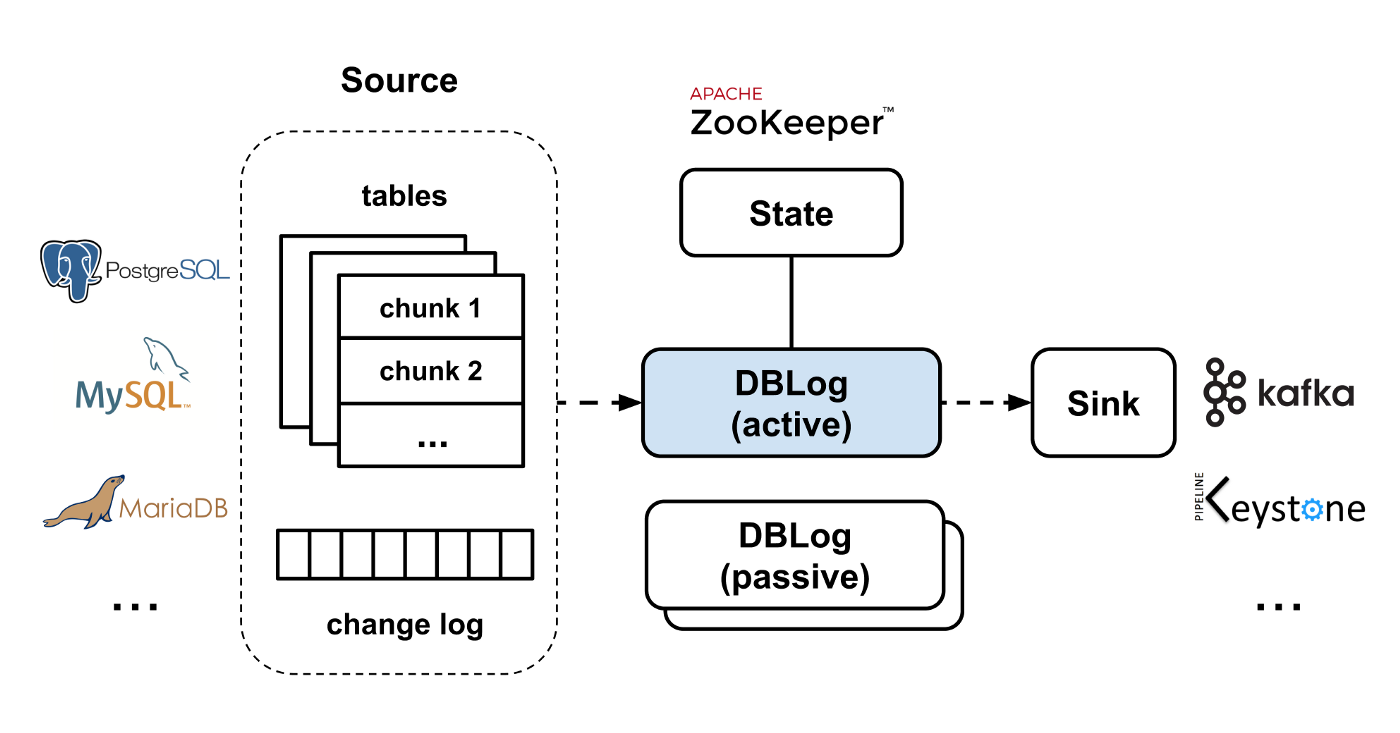}
    \caption{Example of system architecture diagram from a Netflix technical blog post \cite{Andreakis2019a}. Original caption: \enquote{DBLog High Level Architecture}.}
    \label{fig:diagrams-examples}
\end{figure}

To illustrate an example of inconsistency in architectural diagrams, consider the diagram in Figure \ref{fig:diagrams-examples}. Analyzing this diagram, we could formulate several questions. For example, (1) how many services does this system have?; (2) where is this system deployed?; (3) how much does this diagram consistently communicate enough details to represent the same system using other languages or graphical notations, such as UML?

Informally, we analyzed 20 architectural diagrams\footnote{In future work, we must systematically investigate our data set of gray literature architectural diagrams.} from technical blog posts, such as Amazon, LinkedIn, Netflix, and Spotify. For each diagram, at least one of these issues occurs:

\paragraph{\textbf{Issue 1}: Use of boxes associated with different border styles.}
Rectangles or squares are generally associated with different border styles to separate or classify different types of elements. These specifications are not always provided in a caption.

We observe in Figure \ref{fig:diagrams-examples} it is hard to infer how each box is classified, raising questions such as \textit{``Why this is gray, and that is white?"}, or \textit{``Boxes with dashed borders represent a logical group of components or a closer view (zoom in) of a component?"}. Besides that, visual blocking occurs in some boxes in the diagram, as in the representation of the ``chunks" and the ``passive DBLog." This rise to multiple interpretations of the diagram.

\paragraph{\textbf{Issue 2}: Use of arrows.}
It is not clear what each arrow represents in the diagram. For example, arrows represent a flow of data or connections between components.

In software engineering, consistency model validation must ensure that all diagrams correctly express the system requirements \cite{Amor2011}. Some approaches have been proposed to validate model consistency, such as using consistency rules \cite{Ha2003, Amor2011}, algorithmic approach \cite{Litvak2003} or the use of graphs \cite{Mohammadi2014}. However, these approaches validate a UML model from another \cite{Litvak2003,Ha2003,Mohammadi2014}. For example, Mohammadi \textit{et al.} proposed an approach to validate UML component and deployment diagrams \cite{Mohammadi2014}.


The consistency issues can arise because more than one diagram can describe the same aspects of the model. Moreover, even if the models are consistent, there is no guarantee that these models are consistent with the production system. Thus, we propose a new approach to address these issues.

%
%
\section{Approach}
\label{sec:approach}

We propose an approach to improve architectural diagram consistency using system descriptors. Hence, in this section, we briefly present the concepts behind our approach and our hypotheses.
%
%
\subsection{Concepts}
\label{sec:concepts}

According to Ludewig \cite{Ludewig2004}, the architectural diagram is a prescriptive model that specifies the architecture that must be created. A UML deployment diagram is an instance of an architectural diagram.

According to the UML Specification \cite{uml2017omg}, \enquote{\textbf{deployment diagrams} show the configuration of run-time processing elements and the software components, processes, and objects that execute on them}. One of the deployment diagram functions is to map the software architecture to the hardware \cite{Arlow2005}. Deployment diagrams are composed of nodes, communication associations and, where desired, dependency associations between nodes. Besides, run-time components and objects are represented in nodes \cite{Swain2010}.


\textbf{Systems descriptors} are scripts for automation, standardization, and infrastructure management in production environments. 
System descriptors emerge within the context of Infrastructure as Code (IaC), which specifies the definition and set up of the software infrastructure required to run a system by using configuration scripts \cite{Artac2017}. Descriptor providers, such as Docker-Compose, Chef, Puppet, Kubernetes Pods tools, use different languages such as JSON, YAML, or even in specific domain languages (DSL), as Puppet to implement their scripts.
Hence, system descriptors are artifacts that describe a system architecture. 


\textbf{Diagram-as-Code} is an approach to generate diagrams through programming. According to \cite{Mingrammer2020},  diagram-as-code is the design of system architecture diagrams, using a domain-specific language to describe the diagram's elements and relationships. 
This approach has sparked recent interest among software engineering practitioners. A non-exhaustive list of some web articles on the subject is given in \cite{Flaatten2020, Meyer2019, Mingrammer2020, Brown2020a}. 



%
%
\subsection{Hypotheses}
\label{sec:hypotheses}
We propose the use of system descriptors to improve the architectural diagrams consistency, generating them and validating them. Thus, we formulate the following hypotheses:

\begin{hypothesis}{the architectural diagram consistency}
	If an architectural diagram is generated from a valid system descriptor then the diagram is consistent.
	\label{sec:h1}
\end{hypothesis}

We assume that a descriptor file is naturally consistent for two reasons: (1) System descriptor files are written in a formal language, such as YAML (used by Docker-compose and Kubernetes Pods) or Ruby (used by Chef); (2) System descriptor files are automatically processed by a finite state machine. 
If the diagram generated from system descriptors presents each descriptor item, this diagram will also be consistent.
\\
\begin{hypothesis}{the system descriptor validity}
	If a valid system descriptor is generated from an architectural diagram then the diagram is consistent.
	\label{sec:h2}
\end{hypothesis}

We state that if a architectural diagram has accurate enough data to generate a system descriptor that is valid and executable in a tool that processes system descriptor files, then this diagram is consistent. 

\subsection{Transformation function}
\label{sec:Transformation-function}
Let's assume that $S$ is the set of all script instances ($\sigma$) of a system descriptor and $D$ is the set of architectural diagrams ($\delta$). We define the transformation function as $f: S \rightarrow D$, so that:

\begin{equation}  
f(S) = \{\sigma \in S \mid (\exists \delta \in D), f(\sigma) = \delta\} \subset D
\end{equation}
is the image of $f$. Considering that hypotheses 1 and 2 are true, we then assume that the transformation function will have a bijective mapping, so that:
\begin{equation}
   f(\sigma) = \begin{cases}
            \sigma_1, \sigma_2 \in S \mid  \sigma_1 \ne \sigma_2 \Longleftrightarrow f(\sigma_1) \ne f(\sigma_2).\\
            \delta \in D \mid (\exists \sigma \in S)[(\sigma,\delta) \in f].
        \end{cases}
\end{equation}

Hence, if $f$ is a bijector function, then it admits an inverse function $f^{-1}: D \rightarrow S$, such that $f^{-1}(\delta) = \sigma$.

The next section describes a tool that implements the transformation function. This tool was developed and used to explore the hypotheses of this work.

%
%
\section{Case study}
\label{sec:case-study}

We conducted a case study to test our hypotheses. First, we created a meta-model to describe how to transform docker-compose system descriptors into architectural diagrams. Our model defines three top-level entities: \textit{service, volume and network}. The \textit{service} entity contains the settings that we apply to each container within a service. A \textit{volume} describes permanent storage in services, and \textit{network} describes the logical network in a container.

We developed a tool that generates architectural diagrams named
\textbf{Descriptor to Architectural Diagram} or \textbf{DAD}\footnote{\url{https://github.com/jalvesnicacio/descriptors-diagrams}}. In its current version, DAD only recognizes docker-compose as a system descriptor and uses a diagrams-as-code library to generate architectural diagrams \footnote{\url{https://diagrams.mingrammer.com/}}.
Thus, DAD receives a system descriptor file as input, and it generates a architectural diagram and a diagram-as-code script. Listing \ref{lst:diagram-as-code-generated} shows a fragment of the diagram-as-code script automatically generated by the tool. 

\begin{lstlisting}[caption={Diagram-as-code script generated by the DAD tool}, label={lst:diagram-as-code-generated},language=Python]
with DaC("dblog system", direction="TB"):
  with Cluster("mysql service"):
    mysql = Server("mysql")
  with Cluster("dblog service"):
    connect = Server("connect")
  with Cluster("kafka service"):
    kafka = Server("kafka")
  kafka >> zookeeper
  connect - zookeeper
  ...
\end{lstlisting}

\begin{figure}[h]
	\centering
	\includegraphics[width=\columnwidth]{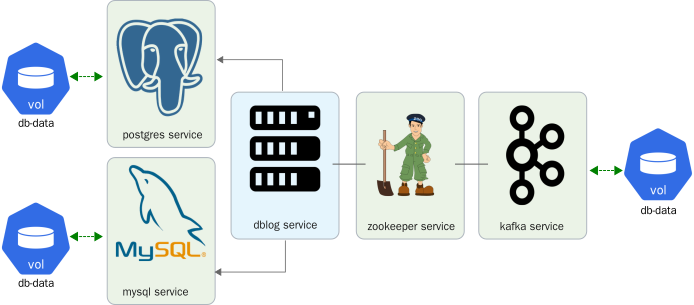}
	\caption{Architectural diagram corresponding to Figure \ref{fig:diagrams-examples}, generated from the DAD tool and modified manually to add some semantic aspects.}
	\label{fig:dac-diagram-neflix}
\end{figure}

We created a docker-compose.yaml file to represents the architectural diagram shown in Fig. \ref{fig:diagrams-examples}. Next, we use DAD to automatically generate the architectural diagram from this docker-compose file, whose fragment is present in Listing \ref{lst:docker-compose}. Finally, we compare the architectural diagram shown in Fig. \ref{fig:diagrams-examples} with an architectural diagram for the same system generated by our tool (Fig. \ref{fig:dac-diagram-neflix}).

\section{Discussion}
\label{sec:discussion}

We compared the architectural diagram from Netflix (Figure \ref{fig:diagrams-examples}) with architectural diagrams generated using DAD (Figure \ref{fig:dac-diagram-neflix}), and we discuss our observation as follows. 

We observed some elements of the original architectural diagram in Figure \ref{fig:diagrams-examples} do not appear in the generated diagram, such as ``State" and ``Sink" boxes.  When these elements are inserted in the architectural diagram, even understanding that they are part of the Zookeeper and Kafka concepts, they could cause misunderstandings because it uses the same symbol for different concepts (for example, ``DBlog" vs ``State"). 

\begin{lstlisting}[caption={Docker-compose.yaml example}, label={lst:docker-compose}]
services:
  mysql:
    image: mysql
  dblog:
    build:
      context: api
      dockerfile: Dockerfile
      container_name: dblog
      depends_on:
         - mysql
         - postgres
...
\end{lstlisting}

As noted in Figure \ref{fig:dac-diagram-neflix}, the diagram that we generated from the system descriptor (docker-compose) contains fewer ambiguous elements than the original diagram. Hence,  the generated diagram is potentially more consistent. 

\begin{tcolorbox}[colframe=gray!50, coltitle=black]
	\textbf{Observation 1: }The generated architectural diagram does not present inconsistent aspects of the original diagram.
\end{tcolorbox}

We also observed in Figure \ref{fig:dac-diagram-neflix} that arrows were represented into an inverted direction concerning Figure \ref{fig:diagrams-examples}. In graph theory, the arrows' orientation refers to a dependency relationship between the two connected elements (asymmetric digraphs concept). Thus, we observe that the relationship between DBLog service and MySQL service could be expressed as \textit{``The DBLog service depends on the MySQL service"}, which was reported in the docker-compose file (see line 12 from Listing \ref{lst:docker-compose}) that we created to be used as input to the DAD tool.

Besides, we note that all docker-compose elements, such as services, volumes, dependency relationships or links, are represented by a uniform and consistent graphical notation in the architectural diagram generated by the DAD tool.
\\
\begin{tcolorbox}[colframe=gray!50, coltitle=black]
	\textbf{Observation 2:} All elements in the Docker-compose description are able to be graphically represented in the generated architectural diagram.
\end{tcolorbox}

Finally, we note that comprehension of architectural diagrams also involves the degree of proximity between the diagram elements. It suggests that the order in which the elements appear in the diagram is important for understanding the diagram, as it expresses the elements' semantic organization. For example, elements representing objects with similar roles should be arranged closer together than those with distinct roles.
\\
\begin{tcolorbox}[colframe=gray!50, coltitle=black]
	\textbf{Observation 3:} The position of an element in the diagram is important for understanding the system.
\end{tcolorbox}

In this version, when the DAD tool generates the architectural diagram, the meta-model considers the order in which the elements appear in the system descriptor's text. It also affects the order in which the elements appear on the diagram. Hence, in Fig. \ref{fig:dac-diagram-neflix}, we adapted manually the diagram generated by the tool to include semantic aspects. We modify the order in which the elements appear in the diagram so that elements that have similar roles remain together, such as database systems. 

\section{Related Work}
Up to our knowledge, this is the first work that proposes an approach to improving architectural diagrams consistency using system descriptors. Related studies are from Paraiso \textit{et al.} \cite{Paraiso2016} and Burco \textit{et al.} \cite{Burco2020}. 

One difference with our work is that we focus on verifying architectural diagrams consistency. In contrast, previous studies aim to validate the system descriptor file still in the initial implementation stages.

%
%

\section{Conclusion}
\label{sec:conclusion}
This paper presents our proposal to apply system descriptors to improve architectural diagrams consistency. We discussed the issues of consistency validation in diagrams,  and we observed the \textit{ad-hoc} system modelling with the current notations does not deal with consistency in architectural diagrams. Hence, we stated two main hypotheses formulated from our observations in real-world architectural diagrams. Further, we presented DAD, a tool that uses Diagram-as-code to create diagrams from Docker-Compose files.

Our case study shows that all Docker-compose description elements can be graphically represented in a generated architectural diagram. Also, the generated architectural diagram does not present inconsistent aspects of the original diagram. Also, we highlight the position of elements in the diagrams is important for understanding the system. 

We did not evaluate whether the diagram-as-code library's notation is sufficient to represent diagrams from descriptors, and it is not the objective of this work to evaluate notations.

Some details of the system descriptor, such as ports and passwords, were not mapped in the generated architectural diagram. Such information describes details that out-of-scope of architectural diagrams. Unmapped elements of a system descriptor can be specified as a subset of the domain of that mapping.

This work presents preliminary research results, and our case study does not prove our hypotheses. However, despite limitations, the case study's observations indicate that the proposed approach is promising and presents a consistent approach for representing system descriptors.
Our case study uses only a single systems descriptor provider: docker-compose. In future work, we intend to carry out studies with others descriptor providers. Also, in its current version, DAD cannot generate a system descriptor from a diagram, which did not evaluate hypothesis 2. As future work, we intend to evolve our tool to evaluate our hypotheses more deeply in a controlled experiment.

\balance
\bibliographystyle{IEEEtran}
\bibliography{IEEEabrv,main}

\vspace{12pt}
\end{document}